\documentclass[12pt]{article}
\usepackage{latexsym,amssymb}
\usepackage{amsmath,amsbsy}
\usepackage[all]{xy}
\usepackage{amsopn,amstext}
\usepackage{cite}
\usepackage{amssymb}
\usepackage{rotating}
\usepackage{amsmath}
\usepackage{amsopn,amstext}
\usepackage{pdflscape}

\allowdisplaybreaks[4]

\renewcommand{\theequation}{\arabic{section}.\arabic{equation}}

\begin{document}

\def\bea{\begin{eqnarray}}
\def\eea{\end{eqnarray}}

\def\no{\nonumber}

\def\lt{\left}
\def\rt{\right}

\baselineskip=20pt

\newcommand{\Title}[1]{{\baselineskip=26pt
   \begin{center} \Large \bf #1 \\ \ \\ \end{center}}}
\newcommand{\Author}{\begin{center}
   \large \bf
Guang-Liang Li${}^{a,b}$, Junpeng Cao${}^{b,c,d,e}\footnote{Corresponding author: junpengcao@iphy.ac.cn}$, Kun Hao${}^{b,f,g}$, Pei Sun${}^{g,h}$, Xiaotian
Xu${}^{b,f,g}$, Tao Yang${}^{b,f,g}$ and Wen-Li
Yang${}^{b,f,g,h}\footnote{Corresponding author:
wlyang@nwu.edu.cn}$
 \end{center}}

\newcommand{\Address}{\begin{center}
${}^a$ Ministry of Education Key Laboratory for Nonequilibrium Synthesis and Modulation of Condensed Matter, School of
Physics, Xi'an Jiaotong University, Xi'an 710049, China\\
${}^b$ Peng Huanwu Center for Fundamental Theory, Xi'an 710127, China\\
${}^c$ Beijing National Laboratory for Condensed Matter Physics, Institute of Physics, Chinese Academy of Sciences, Beijing 100190, China\\
${}^d$ School of Physical Sciences, University of Chinese Academy of Sciences, Beijing 100049, China\\
${}^e$ Songshan Lake Materials Laboratory, Dongguan, Guangdong 523808, China \\
${}^f$ Institute of Modern Physics, Northwest University, Xi'an 710127, China\\
${}^g$ Shaanxi Key Laboratory for Theoretical Physics Frontiers, Xi'an 710127, China\\
${}^h$ School of Physics, Northwest University, Xi'an 710127, China\\
\end{center}}

\Title{Spectrum of the transfer matrices of the  spin chains associated with the $A^{(2)}_3$ Lie algebra}

\Author

\Address
\vspace{0.3truecm}
\begin{abstract}
We study the exact solution of quantum integrable system associated with the $A^{(2)}_3$ twist Lie algebra, where the boundary reflection matrices have
non-diagonal elements thus the $U(1)$ symmetry is broken. With the help of the fusion technique, we obtain the closed recursive relations of the fused transfer matrices. Based on them, together with the asymptotic behaviors and the values at special points, we obtain the eigenvalues and Bethe ansatz equations of the system. We also show that the method is universal and valid for the
periodic boundary condition where the $U(1)$ symmetry is reserved. The results in this paper can be applied to studying the exact solution of the $A^{(2)}_n$-related integrable models with arbitrary $n$.

\vspace{0.5truecm} \noindent {\it PACS:} 75.10.Pq, 02.30.Ik, 71.10.Pm

\end{abstract}
\newpage
\section{Introduction}
\label{intro} \setcounter{equation}{0}

Since the pioneer work of Sklyanin \cite{Sklyanin}, the quantum integrable systems with open boundary conditions draw many attentions.
The open boundary conditions are characterized by the reflection matrices. The integrability of the system requires that the
reflection matrix satisfies the reflection equation. If the reflection matrix is diagonal, the conventional Bethe ansatz methods including the coordinate \cite{c1}
and algebraic \cite{a1,a2,a3} ones can be applied to solve it successfully. However, if the reflection matrix has some non-diagonal elements, the $U(1)$ symmetry is broken and
these traditional methods do not work because of lacking the vacuum/reference state.
Then many interesting methods such as the  q-Qnsager alegebra \cite{q-1, Bas07, Bas10, Bas13}, the separation of variables \cite{Skl95, Fra08, Fra11, Nic12},
the off-diagonal Bethe ansatz (ODBA) \cite{cao13, wang15}, and the modified algebraic Bethe ansatz \cite{Bel13, Bel15, Pim15, Ava15} have been  proposed.

Recently, the study of quantum integrable systems with high ranks
becomes a hot topic due to the many applications in the quantum
field theory, AdS/CFT correspondence in string theory and high
energy physics. The most typical and simple case is the integrable
models associated with $A$-series Lie algebras. The model with
periodic or diagonal open boundary conditions have been studied
extensively \cite{NYRes, NYReshetikhin2, Rafael1, Rafael2, Li1}.
Then the results of the system with non-diagonal boundary
reflections are necessary. The exact solution of $q$-deformed
$su(n+1)$ invariant quantum spin chain, which is connected with
the $A^{(1)}_n$ Lie algebra, has been obtained by using the nested
ODBA \cite{Cao14}. The next task is to study the quantum
integrable models associated with the $A^{(2)}_n$ twist Lie
algebra. For the simplest case, the exact solution of
Izergin-Korepin model \cite{ik}, which is connected with the
$A^{(2)}_2$ Lie algebra, with generic integrable open boundary
condition has been obtained \cite{ik-1}. However, the results with
$n\geq3$ are still missing. We shall note that the generic
integrable boundary reflection of quantum integrable models
related with other twist Lie algebra such as $D^{(2)}_n$ is also
an interesting issue \cite{guan,5-2,5-3,5-4}.

In this paper, we study the exact solution of the $A^{(2)}_3$
model with open boundary condition where the reflection matrices
have  non-diagonal elements. We use the fusion technique
\cite{Kul81,Kul86,Kar79,Kir86,Kir87,Mez1,Mez92}. We find that the
fusion properties of present system are quite different from the
$A^{(1)}_n$ case. In the latter case, only the anti-symmetric
fusion is used. For the present case, the $R$-matrix has two
degenerated points. Based on this fact, we obtain two projectors.
These two projectors give different fused behaviors. With the help
of fused transfer matrices, we find that the fusion processes can
be closed. From the analyzing of polynomials, instead of
constructing the eigenstates, we obtain the eigenvalues of the
system, where the asymptotic behaviors and special points are
used. Then we obtain the energy spectrum of the model Hamiltonian.
In order to show the universality of this method, we also give the
corresponding results of the system with periodic boundary
condition.

The paper is organized as follows. In section 2, we give the description of the model, where the transfer matrix, Hamiltonian, $R$-matrix and reflection matrices are introduced.
In section 3, we study the fusion properties. In section 4, the closed recursive fusion relations among the fused transfer matrices are given.
In section 5, by constructing the inhomogeneous $T-Q$ relations, we obtain the eigenvalues and the corresponding Bethe ansatz equations of the system with non-diagonal boundary reflections.
In section 6, the results associated with the periodic boundary condition are given. The summary of main results and some concluding remarks are presented in section 7. Some detailed calculations
are given in Appendix A.

\section{Associated conserved quantities}
\setcounter{equation}{0}

For the open boundary condition, the one-dimensional quantum integrable systems associated
with the $A^{(2)}_3$ twist Lie algebra is generated by the transfer
matrix $t(u)$
\begin{equation}
t(u)= tr_0 \{ K_0^{+}(u)T_0(u) K^{-}_0(u)\hat{T}_0 (u)\}, \label{trweweu1110}
\end{equation}
where $u$ is the spectral parameter, $tr_0$ means the trace in the four-dimensional auxiliary space $V_0$, $K^{-}_0(u)$ is the reflection matrix at one end and is defined in the auxiliary space $V_0$,
$K_0^{+}(u)$ is the dual one at the other end,
$T_0(u)$ is the monodromy matrix and the $\hat{T}_0(u)$ is the reflecting one. $T_0(u)$ and $\hat{T}_0(u)$ are constructed by the $R$-matrices as \cite{Sklyanin}
\begin{eqnarray}
&& T_0(u)=R_{01}(u-\theta_1)R_{02}(u-\theta_2)\cdots R_{0N}(u-\theta_N), \label{Mon-1}\\
&&\hat{T}_0 (u)=R_{N0}(u+\theta_N)\cdots R_{20}(u+\theta_{2}) R_{10}(u+\theta_1).\label{Tt11}
\end{eqnarray}
Here $\{\theta_j|j=1, \cdots, N\}$ are the inhomogeneous parameters and $N$ is the number of sites.
The subscript $j$ means the four-dimensional quantum space $V_j$. Thus the physical space is $\otimes_{j=1}^N V_j$.
The $R$-matrix defined in the tensor space $V_1\otimes V_2$ is the $16\times 16$ matrix  \cite{5-12}
\begin{eqnarray}
&&R_{12}(u)=
a(u)\sum_{\alpha\neq\alpha'}[e_1]^{\alpha}_{\alpha}\otimes
[e_2]^{\alpha}_{\alpha}
+b(u)\sum_{\alpha\neq\beta,\beta'}[e_1]^{\alpha}_{\alpha}\otimes [e_2]^{\beta}_{\beta}\nonumber\\
&&\qquad+\Big\{e(u)\sum_{\alpha<\beta,\alpha\neq\beta'}
+\bar{e}(u)\sum_{\alpha>\beta,\alpha\neq\beta'}\Big\}[e_1]^{\alpha}_{\beta}\otimes
[e_2]^{\beta}_{\alpha}
+\sum_{\alpha,\beta}a_{\alpha\beta}(u)[e_1]^{\alpha}_{\beta}\otimes
[e_2]^{\alpha'}_{\beta'}, \label{RA2odd}
\end{eqnarray}
where $\alpha, \beta=1, \cdots, 4$, $\alpha'=5-\alpha$,
$\beta'=5-\beta$, $[e_k]^{\alpha}_{\beta}$ is the $4 \times 4$
Weyl basis of the space $V_k$. The matrix elements in Eq.\eqref{RA2odd} are
\begin{eqnarray}
&& a(u)=2\sinh\big(\frac{u}{2}-\eta\big)\cosh\big(\frac{u}{2}-2\eta\big),\quad
b(u)=2\sinh \frac{u}{2}\cosh\big(\frac{u}{2}-2\eta\big),\nonumber\\
&&
e(u)=-2e^{-\frac{u}{2}}\sinh\eta\cosh\big(\frac{u}{2}-2\eta\big),\quad \bar{e}(u)=e^ue(u),\nonumber\\
&& a_{\alpha\beta}(u)=2\sinh\eta e^{\mp\frac{u}{2}} \Big[\mp e^{(\pm
2+\bar{\alpha}-\bar{\beta})\eta}\sinh\frac{u}{2}
-\delta_{\alpha\beta'}\cosh\big(\frac{u}{2}-2\eta\big)\Big],\;\; {\rm if} \;\; \alpha \lessgtr \beta,\nonumber\\
&& a_{\alpha\beta}(u)=2\sinh\frac{u}{2}\cosh\big(\frac{u}{2}-\eta\big), \;\; {\rm if} \;\; \alpha=\beta, \alpha\neq\alpha',
\end{eqnarray}
where $\eta$ is the crossing parameter, $\bar{\alpha}=\alpha+\frac{1}{2}$ if $1\leq\alpha\leq 2$
and $\bar{\alpha}=\alpha-\frac{1}{2}$ if $3\leq \alpha\leq 4$. The $R$-matrix \eqref{RA2odd} has the  properties
\begin{eqnarray}
&&\hspace{-1cm}{\rm unitarity}:\;\; R_{12}(u)R_{21}(-u)=\rho_1(u)\times{\rm id}=a(u)a(-u)\times{\rm id},\nonumber\\
&&\hspace{-1cm}{\rm crossing \; unitarity}:\;\; R_{12}(u)^{t_{1}}{M}_{1}R_{21}(-u+8\eta+2i\pi)^{t_{1}}{M}_{1}^{-1}\no\\
 &&\hspace{2.8cm}=R_{12}(u)^{t_{2}}{M}_{2}^{-1}R_{21}(-u+8\eta+2i\pi)^{t_{2}}{M}_{2}=\rho_1(u-4\eta-i\pi),\no\\
&&\hspace{-1cm}{\rm regularity}:\;\; R_{12}(0)=\rho_1(0)^{\frac{1}{2}}{\cal P}_{12},
\end{eqnarray}
where $M_k$ is the $4\times 4$ diagonal matrix
$M_k=diag(e^{2\eta},1,1,e^{-2\eta})$, ${\cal P}_{12}$ is the
permutation operator with the matrix elements $[{\cal
P}_{12}]^{\alpha\gamma}_{\beta\delta}=\delta_{\alpha\delta}\delta_{\beta\gamma}$,
$t_k$ denotes the transposition in the $k$-th space,
$R_{21}(u)={\cal P}_{12}R _{12}(u){\cal P}_{12}$. Besides, the
$R$-matrix (\ref{RA2odd}) satisfies the Yang-Baxter equation
\begin{eqnarray}
R_{12}(u-v)R_{13}(u)R_{23}(v)=R_{23}(v)R_{13}(u)R_{12}(u-v). \label{20190802-1}
\end{eqnarray}

The integrability of the system requires that the boundary reflection matrix $K^{-}(u)$ satisfies the reflection equation
\begin{equation}
 R_{12}(u-v)K_{1}^-(u)R_{21}(u+v) K^-_{2}(v)=
 K^{-}_{2}(v)R_{12}(u+v)K^{-}_{1}(u)R_{21}(u-v), \label{r1}
 \end{equation}
while $K^{+}(u)$ satisfies the dual one
\begin{eqnarray}
&&R_{12}(-u+v){K}^{+}_{1}(u)M_1^{-1}R_{21}
 (-u-v+8\eta+2i\pi)M_1{K}^{+}_{2}(v)\nonumber\\
&&=K^{+}_{2}(v)M_1R_{12}(-u-v+8\eta+2i\pi)M_1^{-1}
K^{+}_{1}(u)R_{21}(-u+v).
 \label{r2}
 \end{eqnarray}
The general solution of reflection equations \eqref{r1}-\eqref{r2}
for the $A_{n}^{(2)}$ vertex model has been constructed by
Lima-Santos et al \cite{lima-1}, Malara et al \cite{lima-2} and
Nepomechie et al \cite{Rafael2}, where the reflection matrices could have
the non-diagonal elements. Here, we focus on the non-diagonal boundary reflections.
Without losing the generality of our method, we chose \bea
K_k^{-}(u)=\left(\begin{array}{cccc}e^{-u}&0&0&e^{\epsilon}\sinh u \\
    0&\frac{-\sinh(u-\eta)}{\sinh \eta}&0&0\\
    0&0&\frac{-\sinh(u-\eta)}{\sinh \eta}&0\\
    e^{-\epsilon}\frac{\sinh u}{\sinh^2 \eta}&0&0&e^{u}\end{array}\right),\label{K-matrix-1}
\eea
where $\epsilon$ is the boundary parameter at one side.
The dual reflection matrix $K^{+}(u)$ is obtained by the mapping
\begin{equation}
K_k^{+}(u)=M_k K_k^{-}(-u+4\eta+i\pi)|_{\epsilon\rightarrow\,
\epsilon'}, \label{ksk111}
\end{equation}
and $\epsilon'$ is the boundary parameter at the other side.
It is easy to check that  the matrices $K^{-}(u)$ and $K^{+}(u)$ cannot be diagonalized simultaneously for generic values of $\epsilon$ and $\epsilon'$.
Although the $U(1)$ symmetry is broken,
the integrability of the system is still held.

From the Yang-Baxter equation (\ref{20190802-1}), reflection equation (\ref{r1}) and dual one (\ref{r2}), one can
prove \cite{Sklyanin} that the transfer matrices with different spectral parameters
commute with each other, i.e., $[t(u), t(v)]=0$. Thus, expanding $t(u)$ with respect to $u$, all the
coefficients are the conserved quantities.
The Hamiltonian is constructed by taking the
derivative of the logarithm of the transfer matrix
\begin{eqnarray}
H&=&\frac{\partial \ln t(u)}{\partial u}|_{u=0,\{\theta_j\}=0} \nonumber \\
&=& \sum^{N-1}_{j=1}{\cal P}_{j j+1}\left.\frac{\partial R_{j
j+1}(u)}{\partial u}\right|_{u=0}
+\frac{{K^{-}_N}(0)'}{2{K^{-}_N}(0)}+\frac{ tr_0
\{K^{+}_0(0)H_{10}\}}{tr_0 K^{+}_0(0)}+{\rm constant}, \label{hh}
\end{eqnarray}
where $H_{10}={\cal P}_{1 0}\frac{\partial R_{1 0}(u)}{\partial
u}|_{u=0}$. We shall note that because the $R$-matrix
\eqref{RA2odd} reduces to the permutation operator at the point of
$u=0$, the interaction in the bulk is the nearest neighbor one.

\section{Fusion procedure}
\setcounter{equation}{0}

\subsection{Fusion of $R$-matrices}

The next task is to exact diagonal the transfer matrix (\ref{trweweu1110}).
According to the definition, we know that $t(u)$ is an operator-valued polynomial of $e^{u}$ with
degrees $4N+4$, up to an overall factor $e^{-2Nu-2u}$. Thus $t(u)$ can be completely determined by $4N+5$
constraints. In order to obtain these constraints, we adopt  the method of fusion.

It is easy to check that the $R$-matrix (\ref{RA2odd}) degenerates into the projectors at some special points.
For examples, the $R$-matrix degenerates into an one-dimensional projector $P^{(1)}_{12}$ if $u=4\eta+i\pi$, and a
six-dimensional projector $P^{(6)}_{12}$ if $u=2\eta$. These conclusions are achieved by the facts
\bea
R_{12}(4\eta+i\pi)=P^{(1) }_{12}S_{12}^{(1)}, \quad  R_{12}(2\eta)=P^{(6)}_{12}S_{12}^{(6)}, \eea
where $S_{12}^{(1)}$ and $S_{12}^{(6)}$ are the irrelevant constant matrices omitted here,
$P^{(1)}_{12}$ and $P^{(6)}_{12}$ are the projectors
\bea
P^{(1) }_{12}=|\psi_0\rangle\langle\psi_0|, \quad P^{(6)}_{12}=\sum_{i=1}^{6} |{\phi}_i\rangle\langle{\phi}_i|. \label{a1}
\eea
The basis vectors of the related projectors are
\bea &&|\psi_0\rangle=\frac{1}{2\cosh\eta}
(e^{-\eta}|14\rangle+|23\rangle+|32\rangle+e^{\eta}|41\rangle),\no \\
&&|{{\phi}}_1\rangle=\frac{1}{\sqrt{2\cosh
\eta}}(e^{-\frac\eta 2}|12\rangle-e^{\frac\eta 2}|21\rangle),\
|{{\phi}}_2\rangle=\frac{1}{\sqrt{2\cosh
\eta}}(e^{-\frac\eta 2}|13\rangle-e^{\frac\eta 2}|31\rangle),\no\\
&&|{{\phi}}_3\rangle=\frac{1}{\sqrt{2}{\cosh
\eta}}(\sinh\eta|23\rangle+\sinh\eta|32\rangle+|14\rangle-|41\rangle),\no\\
&&|{{\phi}}_4\rangle=\frac{1}{\sqrt{2\cosh
\eta}}(e^{-\frac\eta 2}|23\rangle-e^{\frac\eta 2}|32\rangle),\
|{{\phi}}_5\rangle=\frac{1}{\sqrt{2\cosh
\eta}}(e^{-\frac\eta 2}|24\rangle-e^{\frac\eta
2}|42\rangle),\no\\
&&|{{\phi}}_6\rangle=\frac{1}{\sqrt{2\cosh \eta}}(e^{-\frac\eta
2}|34\rangle-e^{\frac\eta 2}|43\rangle), \eea Exchanging two
spaces $V_{1}$ and $V_{2}$, we obtain $P^{(1)}_{21}$ and
$P^{(6)}_{21}$, where the bases are \bea
|\psi_0\rangle_{|kl\rangle\rightarrow |lk\rangle}, \quad
|{{\phi}}_i\rangle_{\eta\rightarrow-\eta,\,
|kl\rangle\rightarrow|lk\rangle}. \label{1a1} \eea where
$\{|k\rangle, k=1,\cdots, 4\}$ and $\{|l\rangle, l=1,\cdots, 4\}$
are the orthogonal bases of four-dimensional linear space $V_{1}$
and $V_{2}$,respectively.

From the Yang-Baxter equation (\ref{20190802-1}) and using the properties of projector, we obtain
\bea
&&P^{(1) }_{12}R _{23}(u)R _{13}(u+4\eta+i\pi)P^{(1)}_{12}=a(u)c(u+4\eta+i\pi)P^{(1)}_{12},\label{hhhgg-1} \\
&&P^{(1) }_{21}R _{32}(u)R_{31}(u+4\eta+i\pi)P^{(1) }_{21}=a(u)c(u+4\eta+i\pi)P^{(1)}_{21}, \label{hhgg-1}\\
&&P^{(6) }_{12}R _{23}(u)R_{13}(u+2\eta)P^{ (6) }_{12}=\tilde{\rho}_0(u) R_{\langle 12\rangle3}(u+\eta), \label{uf-12} \\
&&P^{(6) }_{21}R _{32}(u)R_{31}(u+2\eta)P^{ (6)
}_{21}=\tilde{\rho}_0(u) R_{3\langle 12\rangle}(u+\eta),
\label{fu-12}
 \eea
where $c(u)=2\sinh \frac u2\cosh(\frac u2-\eta)$, $\tilde{\rho}_0(u)=\sinh\frac{u+\eta}{2}\cosh\frac{u-5\eta}{2}$ and the
subscript $\langle 12\rangle$ denotes the six-dimensional fused space $V_{\langle
12\rangle}=V_{\bar 1}$. From Eqs.\eqref{hhhgg-1}-\eqref{hhgg-1}, we see that the fusion
with one-dimensional projectors gives an one-dimensional vector. From Eqs.\eqref{uf-12}-\eqref{fu-12}, we know that the fusion with
six-dimensional projectors gives a new fused $R$-matrix $R_{\bar{1}2}(u)$, whose matrix elements are given in Appendix A (see (\ref{rsp})-(\ref{A-2}) below).
Moreover, we have checked that $R_{\bar{1}2}(u)$ also satisfies the properties
\begin{eqnarray}
&&\hspace{-1cm}{\rm unitarity}: \;\; R_{\bar{1}2}(u) R_{2\bar{1}}(-u)\times{\rm id}=\rho_2(u)=a_1(u)a_1(-u)\times{\rm id},\nonumber\\
&&\hspace{-1cm}{\rm crossing \; unitarity}: \;\; R_{\bar{1}2}(u)^{t_{\bar{1}}}{\bar{M}}_{\bar 1}R_{2\bar{1}}(-u+8\eta+2i\pi)^{t_{\bar{1}}}{\bar{M}}_{\bar{1}}^{-1}\no\\
 &&\hspace{2cm} =R_{\bar{1}2}(u)^{t_{2}}{M}_{2}^{-1}R_{2\bar{1}}(-u+8\eta+2i\pi)^{t_{2}}{M}_{2}=\rho_2(u-4\eta-i\pi), \no \\
&&\hspace{-1cm}{\rm periodicity}: \;\;
R_{\bar{1}2}(u+i\pi)=-{\bar{V}}_{\bar{1}}R_{\bar{1}2}(u){\bar{V}}_{\bar{1}}^{-1},\label{nrp}
\end{eqnarray}
where $a_1(u)=2\sinh(u-3\eta)$, $\bar{M}_{\bar{1}}$ is the
diagonal matrix $\bar{M}_{\bar{1}}=P^{ (6) }_{12}M_1M_2P^{
(6)}_{12}=diag(e^{2\eta}, e^{2\eta}, $ $ 1, 1, e^{-2\eta},
e^{-2\eta})$ and ${\bar{V}}_{\bar{1}}$ is a $6\times 6$ matrix
with the form of
\begin{eqnarray}
{\bar{V}}_{\bar{1}}=\lt(\begin{array}{cccccc}1&&&&&\\
&-1&&&&\\&&&1&&\\&&1&&&\\&&&&-1\\
&&&&&1
\end{array}\rt).
\end{eqnarray}
The above properties are very useful later for us to derive some important polynomial properties of the associated transfer matrices $\bar{t}(u)$ given by (\ref{transfer-open})
and $\bar{t}^{(p)}(u)$ given by (\ref{1117-1}).

It is remarked that the fused $R$-matrix $R_{\bar{1}2}(u)$ becomes a $4\times 4$ matrix at the point of $u=3\eta$
\bea
R_{\bar{1}2}(3\eta)=P_{\bar{1}2}^{(4)}S_{\bar{1}2}^{(4)}, \quad P_{\bar{1}2}^{(4)}=\sum_{i=1}^{4}
|\varphi_i\rangle\langle\varphi_i|, \eea
where $S_{\bar{1}2}^{(4)}$ is an irrelevant constant matrix omitted here, and $P_{\bar{1}2}^{(4)}$ is a 4-dimensional projector with the basis
vectors
 \bea
&&|\varphi_1\rangle=\frac{1}{\sqrt{2\cosh\eta+e^{3\eta}}}(\sqrt{\cosh\eta}|13\rangle
-\sqrt{\cosh\eta}|22\rangle-e^{\frac{3{\eta}}{2}}|41\rangle),\no\\
&&|\varphi_2\rangle=\frac{1}{\sqrt{1+2\cosh
2\eta}}(e^{-\frac{\eta}{2}}\sqrt{\cosh\eta}|14\rangle-\cosh\eta|32\rangle
-\sinh\eta|42\rangle+e^{\frac{\eta}{2}}\sqrt{\cosh\eta}|51\rangle),\no\\
&&|\varphi_3\rangle=\frac{1}{\sqrt{1+2\cosh
2\eta}}(e^{-\frac{\eta}{2}}\sqrt{\cosh\eta}|24\rangle-\cosh\eta|33\rangle
+\sinh\eta|43\rangle+e^{\frac{\eta}{2}}\sqrt{\cosh\eta}|61\rangle),\no\\
&&|\varphi_4\rangle=
\frac{1}{\sqrt{2\cosh\eta+e^{-3\eta}}}(\sqrt{\cosh\eta}|62\rangle
-\sqrt{\cosh\eta}|53\rangle+e^{-\frac{3{\eta}}{2}}|44\rangle).
\eea  Exchanging the two spaces $V_{\bar{1}}$ and $V_{2}$, we
deduce another 4-dimensional projector $P_{2\bar{1}}^{(4)}$ with
the bases $ |{\varphi}_i\rangle_{\eta\rightarrow
-\eta,|kl\rangle\rightarrow |lk\rangle}$, where $\{|k\rangle,
k=1,\cdots, 6\}$ and $\{|l\rangle, l=1,\cdots, 4\}$ are the
orthogonal bases of six-dimensional linear space $V_{\bar{1}}$ and
four-dimensional linear space $V_{2}$,respectively. Starting from
the Yang-Baxter equation
\begin{eqnarray}
R_{\bar 12}(u-v)R_{\bar 13}(u)R_{23}(v)=R_{23}(v)R_{\bar 13}(u)R_{\bar 12}(u-v), \label{201908102-1}
\end{eqnarray}
and using the properties of projector, we have
\bea &&P^{ (4)}_{\bar{1}2}R _{23}(u)R_{\bar{1}3}(u+3\eta)P^{(4) }_{\bar{1}2}=\tilde{\rho}_1(u){S}_{\langle
\bar{1}2\rangle}R_{\langle \bar{1}2\rangle 3}(u+2\eta+i\pi){S}_{\langle \bar{1}2\rangle}^{-1}, \label{sv-2-1} \\
&&P^{ (4) }_{2\bar{1}}R _{32}(u)R_{3\bar{1}}(u+3\eta)P^{ (4)}_{2\bar{1}}=\tilde{\rho}_1(u){S}_{\langle
\bar{1}2\rangle}R _{3\langle\bar{1}2\rangle}(u+2\eta+i\pi){S}_{\langle
\bar{1}2\rangle}^{-1}, \label{sv-2}
 \eea
where $\tilde{\rho}_1(u)=-4\sinh(\frac u2+\eta)\cosh(\frac
u2-2\eta)$, the subscript $\langle \bar{1}2\rangle$ denotes the
fused four-dimensional space $V_{\langle \bar{1}2\rangle}$, and
${S}_{\langle \bar{1}2\rangle}$ is a diagonal matrix
\begin{eqnarray}
&&{S}_{\langle
\bar{1}2\rangle}=diag\lt(-e^{-\frac{\eta}{2}}\frac{\sinh\eta}{\sinh
3\eta}s(\eta), 1, -1, e^{\frac{\eta}{2}}\frac{\sinh\eta}{\sinh
3\eta}s(-\eta)
\rt), \nonumber \\[4pt]
&&s(\eta)= \sqrt{(1+2\cosh 2\eta)(e^{3\eta}+2\cosh\eta)}.\label{vbar}
\end{eqnarray}
From Eqs.\eqref{sv-2-1} and \eqref{sv-2}, we see that the fused $R$-matrices $R_{\langle \bar{1}2\rangle 3}(u)$ and $R_{3\langle \bar{1}2\rangle}(u)$ differ from the fundamental ones only by a similar transformation up to a constant.
By introducing the one-to-one correspondence, we can map the fused space $V_{\langle \bar{1}2\rangle}$ into $V_1$. Then
the fused $R$-matrix $R_{\langle \bar{1}2\rangle3}(u)$ becomes the fundamental $R$-matrix $R_{13}(u)$ given by (\ref{RA2odd}).
Then we conclude that the fusion processes of $R$-matrices are closed.

\subsection{Fusion of monodromy matrices}

From the fused $R$-matrices \eqref{uf-12}-\eqref{fu-12}, we construct the fused monodromy matrices
\bea
&&T_{\bar{0}}(u)=R_{\bar{0}1}(u-\theta_1)R_{\bar{0}2}(u-\theta_2)\cdots R_{\bar{0}N}(u-\theta_N), \label{M2on1-2} \\
&&\hat{T}_{\bar{0}}(u)=R_{N{\bar{0}}}(u+\theta_N)\cdots R_{2 {\bar{0}}}(u+\theta_{2}) R_{1 {\bar{0}}}(u+\theta_1).\label{M2on-2}
\eea
We should note that the fusions are taken in the auxiliary space, thus all the quantum spaces of $T_{0}(u)$, $\hat T_{0}(u)$, $T_{\bar 0}(u)$ and $\hat T_{\bar{0}}(u)$ are the same.

From the Yang-Baxter equations (\ref{20190802-1}) and \eqref{201908102-1}, we can prove that the monodromy matrices satisfy the Yang-Baxter relations
\bea
&& R_{12}(u-v) T_1(u) T_2(v) = T_2(v) T_1(u) R_{12}(u-v), \label{ybta2o} \\
&&R_{1\bar{2}}(u-v) T_1(u)T_{\bar{2}}(v) = T_{\bar{2}}(v)T_1(u)
R_{1\bar{2}}(u-v).  \label{ya1ng-1} \eea By using the fusion
identities (\ref{hhhgg-1}), (\ref{uf-12}) and (\ref{sv-2-1}), we
obtain \bea &&P^{(1)}_{21}T_1(u)T_2(u+4\eta+i\pi)P^{(1)}_{21}=
P^{(1)}_{21}\prod_{j=1}^N
a(u-\theta_j)c(u-\theta_j+4\eta+i\pi)\times {\rm id}, \no \\
&&P^{(6)}_{12}T_2(u)T_1(u+2\eta)P^{(6)}_{12}=\prod_{j=1}^N \tilde{\rho}_0(u-\theta_j)T_{\langle
12\rangle}(u+\eta),\no \\
&&P_{\bar{1}2}^{(4)}T_2(u)T_{\bar{1}}(u+3\eta)P_{\bar{1}2}^{(4)}=
\prod_{j=1}^N \tilde{\rho}_1(u-\theta_j){S}_{\langle
\bar{1}2\rangle}T_{\langle
\bar{1}2\rangle}(u+2\eta+i\pi){S}_{\langle\bar{1}2\rangle}^{-1}.\label{98opr-11}
 \eea

The reflecting monodromy matrices satisfy the Yang-Baxter relations
\bea
&& R_{ 21}(u-v) \hat T_{1}(u) \hat T_2(v)=\hat  T_2(v) \hat T_{ 1}(u) R_{21}(u-v), \label{haishi0}\\
&&R_{1\bar{2}}(u-v)  \hat T_{\bar{2}}(v)\hat T_{1}(u)=\hat T_{1}(u)\hat  T_{\bar{2}}(v)  R_{1\bar{2}}(u-v). \label{haish1i01}
\eea
From Eqs.(\ref{hhgg-1}), (\ref{fu-12}) and (\ref{sv-2}),
we obtain the fusion identities among the reflecting monodromy matrices \bea
&&P^{(1)}_{12}\hat{T}_1(u)\hat{T}_2(u+4\eta+i\pi)P^{(1)}_{12}=P^{(1) }_{12}\prod_{j=1}^N
a(u+\theta_j)c(u+\theta_j+4\eta+i\pi)\times {\rm id}, \no \\
&&P^{(6)}_{21}\hat{T}_2(u)\hat{T}_1(u+2\eta)P^{(6)}_{21}=\prod_{j=1}^N \tilde{\rho}_1(u+\theta_j)\hat{T}_{\langle
{1}2\rangle}(u+\eta),\no \\
&&P_{2\bar{1}}^{(4)}\hat{T}_2(u)\hat{T}_{\bar{1}}(u+3\eta)P_{2\bar{1}}^{(4)}=
\prod_{j=1}^N \tilde{\rho}_1(u+\theta_j){S}_{\langle
\bar{1}2\rangle}\hat{T}_{\langle
\bar{1}2\rangle}(u+2\eta+i\pi){S}_{\langle\bar{1}2\rangle}^{-1}.\label{opr-1012}
 \eea

\subsection{Fusion of reflection matrices}

Using the fusion technique\cite{Mez1,Mez92}, now, we need to
connect the fusions of monodromy matrices and those of the
reflecting ones, which gives the fusion behavior of the reflection
matrices. We first define the fused transfer matrix \bea
\bar{t}(u)=tr_{\bar{0}}\{ K^{+}_{\bar{0}}(u)
T_{\bar{0}}(u)K^{-}_{\bar{0}}(u)\hat{T}_{\bar{0}}(u)\},\label{transfer-open}
\eea where the trace is taken in the fused auxiliary space
$\bar{0}$ and $K^{\pm}_{\bar{0}}(u)$ are the fused reflection
matrices. Then, we calculate the quantities \bea &&
t(u)t(u+\Delta)=\rho^{-1}_1(2u+\Delta-4\eta-i\pi)tr_{12}\{{K}^{+}_{2}(u+\Delta)M_2^{-1}
R_{12}(-2u+8\eta+2i\pi-\Delta) \no\\[4pt]
&&\quad\quad\times M_2{K}_1^{+}(u)T_1 (u)
T_{2}(u+\Delta)K^{-}_1(u)R_{21}(2u+\Delta)K^{-}_{2}(u+\Delta)\hat{T}_1
(u)\hat{T}_{2}(u+\Delta)\},\label{tt-1}  \\[4pt]
&&t(u)\bar{t}(u+\Delta)=\rho^{-1}_2(2u+\Delta-4\eta-i\pi)tr_{1\bar{2}}
\{{K}^{+}_{\bar{2}}(u+\Delta)\bar{M}_{\bar{2}}^{-1} R_{1\bar{2}}(-2u+8\eta+2i\pi-\Delta)\no\\[4pt]
&&\quad\quad\times
\bar{M}_{\bar{2}}{K}_1^{+}(u)T_1 (u)
T_{\bar{2}}(u+\Delta)K^{-}_1(u)R_{\bar{2}1}(2u+\Delta)K^{-}_{\bar{2}}(u+\Delta)\hat{T}_1
(u)\hat{T}_{\bar{2}}(u+\Delta)\}, \label{tt-2}  \eea where
$\Delta$ is the shift of spectral parameter.
From the fusion of monodromy matrices, we know that
$\Delta$ should be chosen as $4\eta+i\pi$, $2\eta$ in Eq.(\ref{tt-1}) and as $3\eta$ in (\ref{tt-2}), which gives the
fusion relations of reflection matrices.

The $\Delta=4\eta+i\pi$ in Eq.(\ref{tt-1}) corresponds to the fusion with one-dimensional projectors.
According to Eq.(\ref{tt-1}) and using the reflection equations \eqref{r1}-\eqref{r2} and the properties of projector, we obtain
\bea && P_{21}^{(1)}K_{1}^{-}(u)R_{21}(2u+4\eta+i\pi)K_{2}^{-}(u+4\eta+i\pi)P_{12}^{(1)}\no\\
&&\qquad \qquad =\frac{1}{\sinh^2\eta}\sinh(u+4\eta)\sinh(2u+2\eta)\sinh(u-\eta) P_{12}^{(1)}, \label{8005-1}\\
&& P_{12}^{(1)}{K}_{2}^{+}(u+4\eta+i\pi)M_1R_{12}(-2u+4\eta+i\pi)M_1^{-1}
{K}_{1}^{+}(u)P_{21}^{\rm(1)}\no\\
&& \qquad\qquad
=-\frac{1}{\sinh^2\eta}\sinh(u-4\eta)\sinh(2u-2\eta)\sinh(u+\eta)P_{21}^{(1)}. \label{0805-1} \eea
We shall remark that the inserted $R$-matrices with fixed spectral parameters in Eqs.\eqref{8005-1} and \eqref{0805-1} is to reserve the integrability of the system.
The fused results are the one-dimensional vectors.

The $\Delta=2\eta$ in Eq.(\ref{tt-1}) corresponds to the fusion with six-dimensional projectors. From Eq.(\ref{tt-1}), we obtain
that the fused reflection matrices should be constructed as
\bea
\hspace{-1.2truecm}&&\hspace{-1.2truecm}P_{12}^{(6)}K^{-}_2(u)R_{12}(2u+2\eta)K^{-}_1(u+2\eta)P_{21}^{(6)}\no\\
    &&\quad\quad =\frac{2}{\sinh\eta}\cosh(u-\eta)\sinh(u-\eta)\sinh(u+\eta)\sinh(u+2\eta)K^{-}_{\langle 12\rangle}(u+\eta), \label{8005-2} \\
\hspace{-1.2truecm}&&\hspace{-1.2truecm}P^{(6)}_{21}{K}^{+}_1(u+2\eta)\bar{M}_1^{-1}R_{21}(-2u+6\eta)\bar{M}_1{K}^{+}_2(u)P^{(6)}_{12}\no\\
    &&\quad\quad  =-\frac{2}{\sinh\eta}\cosh(u\hspace{-0.02truecm}-\hspace{-0.02truecm}\eta)
    \sinh(u\hspace{-0.02truecm}-\hspace{-0.02truecm}\eta)
    \sinh(u\hspace{-0.02truecm}-\hspace{-0.02truecm}3\eta)
    \sinh(u\hspace{-0.02truecm}-\hspace{-0.02truecm}4\eta)
    {K}^{+}_{\langle 12\rangle}(u\hspace{-0.02truecm}+\hspace{-0.02truecm}\eta), \label{0805-2} \eea
where $K^{-}_{\langle 12\rangle}(u)$ is the $6\times 6$ fused reflection matrix defined in the fused space $V_{\langle 12\rangle}$
with the matrix form of
\bea
K_{\langle 12\rangle}^{-}(u)=\left(\begin{array}{cccccc}0 &0&0&0&e^{\epsilon}&0\\[6pt]
    0&0&0&0&0&e^{\epsilon}\\[6pt]
    0&0&-\frac{1}{\sinh\eta}&0&0&0\\[6pt]
    0&0&0&\frac{1}{\sinh\eta}&0&0\\[6pt]
    \frac{e^{-\epsilon}}{\sinh^2\eta}&0&0&0&0 &0\\[6pt]
0&\frac{e^{-\epsilon}}{\sinh^2\eta}&0&0&0&0
\end{array}\right),
\label{K-matrix-VV}\eea
and ${K}^{+}_{\langle 12\rangle}(u)$ is the dual one
\bea
{K}^{+}_{\langle 12\rangle}(u)=\bar{M}_{\langle 12\rangle}K^{-}_{\langle 12\rangle}(-u+4\eta+i\pi)|_{\epsilon\rightarrow
   \epsilon'}. \eea
With the definition $V_{\bar 1}=V_{\langle 12\rangle}$, the fused reflection matrices satisfy the reflection equations \bea
&& R_{\bar{1}2}(u-v){K^{-}_{ \bar{1}}}(u)R_{2\bar{1}}(u+v) K^{-}_{2}(v)= K^{-}_{2}(v)R_{\bar{1}2}(u+v)K^{-}_{\bar{1}}(u)R_{2\bar{1}}(u-v),  \label{1r2} \\
&&R_{\bar{1}2}(-u+v){K}^{+}_{\bar{1}}(u)\bar{M}_{\bar{1}}^{-1}R_{2\bar{1}} (-u-v+8\eta+2i\pi)\bar{M}_{\bar{1}}{K}^{+}_{2}(v)\nonumber\\
&&\qquad\qquad={K}^{+}_{2}(v)\bar{M}_{\bar{1}}R_{\bar{1}2}(-u-v+8\eta+2i\pi)\bar{M}_{\bar{1}}^{-1}{K}^{+}_{\bar{1}}(u)R_{2\bar{1}}(-u+v). \label{1dr3}
 \eea

The $\Delta=3\eta$ in Eq.(\ref{tt-2}) corresponds to the fusion
with four-dimensional projectors. According to Eq.(\ref{tt-2}),
the fusion of reflection matrices are \bea  &&
P_{\bar{1}2}^{(4)}K_2^{-}(u)R_{\bar{1}2}(2u+3\eta)K_{\bar{1}}^{-}(u+3\eta)P_{2\bar{1}}^{({4})}
\no\\&&\qquad\qquad=\frac{4}{\sinh\eta}\cosh(u)\sinh(u-\eta){S}_{\langle
\bar{1}2\rangle} K_{\langle \bar{1}2\rangle
}^{-}(u+2\eta+i\pi){S}_{\langle
\bar{1}2\rangle}^{-1}, \label{20805-31} \\
&&P_{2\bar{1}}^{(4)}{K}_{\bar{1}}^{+}(u+3\eta)\bar{M}_{\bar{1}}^{-1}R_{2\bar{1}}(-2u+5\eta)\bar{M}_{\bar{1}}{K}_2^{+}(u)P_{\bar{1}2}^{(4)}
\no\\&&\qquad\qquad=-\frac{4}{\sinh\eta}
\cosh(u-\eta)\sinh(u-4\eta){S}_{\langle
\bar{1}2\rangle}{K}^{+}_{\langle
\bar{1}2\rangle}(u+2\eta+i\pi){S}_{\langle \bar{1}2\rangle}^{-1}.
\label{0805-31} \eea With the same one-to-one correspondence as
used in the fusion of $R$-matrices, the fused  reflection matrices
$K_{\langle \bar{1}2\rangle }^{\pm}(u)$ become the original ones
given by Eqs.(\ref{K-matrix-1})-(\ref{ksk111}). Thus the fusion
processes of reflection matrices are also closed.

The fusion does not break the integrability. From the fused Yang-Baxter equation \eqref{201908102-1} and the fused reflection equations \eqref{1r2}-\eqref{1dr3},
one can prove
that the transfer matrices $ t(u)$ and $ \bar{t}(u)$
commutate with each other, i.e., \bea [t(u),\,
\bar{t}(u)]=0. \eea Thus $t(u)$ and $\bar{t}(u)$ have common eigenstates.

\section{Closed operators identities}
\setcounter{equation}{0}

From the definitions of $R$-matrices (\ref{RA2odd}), (\ref{nrp}) and reflection
matrices (\ref{K-matrix-1}), (\ref{K-matrix-VV}), we know that the
$t(u)$ (resp. $\bar{t}(u)$) is an operator-valued polynomial of $e^{u}$
with degree $4N+4$ up to an overall factors $e^{-2Nu-2u}$ ( an operator-valued  polynomial of $e^{2u}$ with degree $2N$ up to an overall factor $e^{-2Nu}$ respectively). Denote the eigenvalues
of $t(u)$ and $\bar{t}(u)$ acting on a common eigenstate as
$\Lambda(u)$ and $\bar{\Lambda}(u)$, respectively. Then the
eigenvalue $\Lambda(u)$ (resp. $\bar{\Lambda}(u)$) is a
polynomial of $e^{u}$ with degree $4N+4$ (is a
polynomial of $e^{2u}$ with degree $2N$) up to an overall known factor.
Therefore, $\Lambda(u)$ and $\bar{\Lambda}(u)$ can be completely
determined by the values of them at $6N+6$ points. The next task
is to find these complete constraints.

From Eqs.(\ref{tt-1})-(\ref{tt-2}), we see that for arbitrary values of spectral parameter $u$, the fusion relations among the transfer matrices $t(u)$ and $\bar{t}(u)$ are not closed.
However, we find that at the inhomogeneous points $\{\theta_j\}$, the fusions of $t(u)$ and $\bar{t}(u)$ can be closed. The detailed derivation is as follows. From the Yang-Baxter relation (\ref{ybta2o}) at the points of $\{u=\theta_j$, $v=\{\theta_j+4\eta+i\pi, \theta_j+2\eta\}\}$,
(\ref{ya1ng-1}) at the points of $\{u=\theta_j$, $v=\theta_j+3\eta\}$
and using the properties of projectors, we obtain
\bea &&T_1(\theta_j)T_2(\theta_j+4\eta+i\pi)=P^{(1) }_{21}T_1(\theta_j)T_2(\theta_j+4\eta+i\pi),  \no \\
&&T_2(\theta_j)T_1(\theta_j+2\eta)=P^{(6) }_{12}\,T_2(\theta_j)T_1(\theta_j+2\eta),\no \\
&&T_2(\theta_j)T_{\bar{1}}(\theta_j+3\eta)=P_{\bar{1}2}^{(4)}T_2(\theta_j)T_{\bar{1}}(\theta_j+3\eta),\quad
j=1,\cdots,N. \label{98opr-1} \eea
We see that we can obtain three projectors by the suitable choices of spectral parameters in the monodromy matrices.
The role of introducing inhomogeneous parameters $\{\theta_j\}$ is to generate the projectors.
The generated projectors allow us to taken the fusion, which
is valid for arbitrary $u$ and the only requirement is the shift $\Delta$.
Substituting Eq.\eqref{98opr-1} into (\ref{tt-1})-(\ref{tt-2}) with $u=\theta_j$, we obtain one set of fusion relations between $t(u)$ and $\bar{t}(u)$.
The Yang-Baxter relation (\ref{haishi0}) at the points of $\{u=-\theta_j$, $v=\{-\theta_j+4\eta+i\pi, -\theta_j+2\eta\}\}$ and (\ref{haish1i01}) at the points of
$\{u=-\theta_j$, $v=-\theta_j+3\eta\}$ give
\bea
&&\hat{T}_1(-\theta_j)\hat{T}_2(-\theta_j+4\eta+i\pi)=P^{(1) }_{12}\hat{T}_1(-\theta_j)\hat{T}_2(-\theta_j+4\eta+i\pi), \no \\
&&\hat{T}_2(-\theta_j)\hat{T}_1(-\theta_j+2\eta)=P_{21}^{(6) }\hat{T}_2(-\theta_j)\hat{T}_1(-\theta_j+2\eta),\no \\
&&\hat{T}_2(-\theta_j)\hat{T}_{\bar{1}}(-\theta_j+3\eta)=P_{2\bar{1}}^{(4)}\hat{T}_2(-\theta_j)\hat{T}_{\bar{1}}(-\theta_j+3\eta), \quad j=1,\cdots, N. \label{op22r-1} \eea
We see that three projectors can also be generated in this situation.
Substituting Eq.\eqref{op22r-1} into (\ref{tt-1})-(\ref{tt-2}) with $u=-\theta_j$, we obtain another set of fusion relations between $t(u)$ and $\bar{t}(u)$.

Now, we are ready to seek the closed fusion relations among the transfer matrices.
Substituting Eqs.(\ref{98opr-11}), (\ref{opr-1012}), (\ref{8005-1})-(\ref{0805-2}), (\ref{20805-31})-(\ref{0805-31}), (\ref{98opr-1})-(\ref{op22r-1}) into Eq.(\ref{tt-1})
and considering the cases of $\{u=\pm\theta_j, \Delta=4\eta+i\pi\}$ and $\{u=\pm\theta_j, \Delta=2\eta\}$,
and into Eq.(\ref{tt-2}) with $\{u=\pm\theta_j, \Delta=3\eta\}$, we arrive at \bea &&
\hspace{-0.8cm}t(\pm\theta_j)t(\pm\theta_j+4\eta+i\pi)=\frac{\sinh(\pm2\theta_j-2\eta)
\sinh(\pm2\theta_j+2\eta)\sinh(\pm\theta_j-4\eta)
\sinh(\pm\theta_j+4\eta) }
{4\sinh^4\eta\cosh(\pm\theta_j-2\eta)\cosh(\pm\theta_j+2\eta)}\no\\
&&\hspace{-0.8cm}\quad\quad\times \prod_{l=1}^N
a(\pm\theta_j-\theta_l)c(\pm\theta_j-\theta_l+4\eta+i\pi)
a(\pm\theta_j+\theta_l)c(\pm\theta_j+\theta_l+4\eta+i\pi)\times {\rm id},\no  \\
&&\hspace{-0.8cm}
t(\pm\theta_j)t(\pm\theta_j+2\eta)=\frac{\sinh^2(\pm2\theta_j-2\eta)
\sinh(\pm\theta_j+2\eta)\sinh(\pm\theta_j-4\eta)}
{4\sinh^2\eta\cosh(\pm\theta_j)\cosh(\pm\theta_j-2\eta)}\no\\
&&\hspace{-0.8cm}\quad\quad\times\prod_{l=1}^N
\tilde{\rho}_0(\pm\theta_j-\theta_l)\tilde{\rho}_0(\pm\theta_j+\theta_l)
 \bar{t}(\pm\theta_j+\eta),\no \\
 &&\hspace{-0.8cm}t(\pm\theta_j)\bar{t}(\pm\theta_j+3\eta)=\frac{2\cosh(\pm\theta_j)
\sinh(\pm2\theta_j-2\eta)\sinh(\pm\theta_j-4\eta)}
{\sinh^2\eta\sinh(\pm2\theta_j+2\eta)\sinh(\pm2\theta_j-4\eta)}\no\\
&&\hspace{-0.8cm}\quad\quad\times\prod_{l=1}^N
\tilde{\rho}_1(\pm\theta_j-\theta_l)\tilde{\rho}_1(\pm\theta_j+\theta_l)
 t(\pm\theta_j+2\eta+i\pi), \quad j=1,\cdots, N. \label{Op-Product-Periodic-41} \eea
We shall note that the recursive equations \eqref{Op-Product-Periodic-41} are closed, which
give the $6N$ constraints of $t(u)$ and $\bar t(u)$ at the inhomogeneous points.

Besides, from the direct calculation, we also obtain the values of $t(u)$ and $\bar t(u)$
at some special points
\bea &&t(0)=4\cosh^2\eta\prod_{j=1}^N\rho_1
(-\theta_j)\times {\rm id},\quad
t(4\eta+i\pi)=4\cosh^2\eta\prod_{j=1}^N\rho_1 (-\theta_j)\times
{\rm id},\no\\
&&t(2\eta)=\frac{\cosh^2 \eta}{2\cosh 2\eta(1+2\cosh
2\eta)}\bar{t}(\eta). \label{2Op} \eea In the derivation, we have
used the relations \bea&&
tr[{K}^+(0)K^-(0)]=4\cosh^2\eta,\quad  tr[{K}^+(4\eta+i\pi)K^-(4\eta+i\pi)]=4\cosh^2\eta,\no\\[4pt]
&&tr_1\{M_2^{-1}R_{12}(6\eta+2i\pi)M_2
{K}^+_1(0)R_{21}(2\eta)\}=4\cosh 2\eta(1+2\cosh 2\eta)\sinh^2
2\eta\times{\rm id}.\no\eea The asymptotic behaviors of $t(u)$ and
$\bar t(u)$ read \bea t(u)|_{u\rightarrow
\pm\infty}=Q_{\pm}e^{\pm(2N+2)u}+\cdots, \quad
\bar{t}(u)|_{u\rightarrow \pm\infty}=\bar
Q_{\pm}e^{\pm2Nu}+\cdots, \label{1Op} \eea where $Q_{\pm}$ and
$\bar Q_{\pm}$ are the conserved quantities \bea &&Q_+=\frac
{1}{4\sinh^2
\eta}\Big\{e^{\epsilon'-\epsilon}e^{-2\eta}[T_+]^4_4[\hat{T}_+]^1_1
+e^{\epsilon-\epsilon'}e^{-6\eta}[T_+]^1_1[\hat{T}_+]^4_4\no\\
&&\hspace{10mm}+e^{-4\eta}[T_+]^2_2[\hat{T}_+]^2_2+
e^{-4\eta}[T_+]^3_3[\hat{T}_+]^3_3\Big\},\no\\
&&Q_-=\frac {1}{4\sinh^2
\eta}\Big\{e^{\epsilon'-\epsilon}e^{6\eta}[T_-]^4_4[\hat{T}_-]^1_1
+e^{\epsilon-\epsilon'}e^{2\eta}[T_-]^1_1[\hat{T}_-]^4_4\no\\
&&\hspace{10mm}+e^{4\eta}[T_-]^2_2[\hat{T}_-]^2_2+
e^{4\eta}[T_-]^3_3[\hat{T}_-]^3_3\Big\},\no\\
&&\bar Q_+=\frac
{1}{\sinh^2\eta}\Big\{e^{\epsilon'-\epsilon}e^{2\eta}\Big([{\bar
T}_+]^5_5[\hat{{\bar T}}_+]^1_1
+[{\bar T}_+]^6_6[\hat{{\bar T}}_+]^2_2\Big)\no\\
&&\hspace{10mm}+[{\bar T}_+]^3_3[\hat{{\bar T}}_+]^3_3 +[{\bar
T}_+]^4_4[\hat{{\bar
T}}_+]^4_4+e^{\epsilon-\epsilon'}e^{-2\eta}\Big([{\bar
T}_+]^1_1[\hat{{\bar T}}_+]^5_5
+[{\bar T}_+]^2_2[\hat{{\bar T}}_+]^6_6\Big)\Big\},\no\\
&&\bar Q_-=\frac
{1}{\sinh^2\eta}\Big\{e^{\epsilon'-\epsilon}e^{2\eta}\Big([{\bar
T}_-]^5_5[\hat{{\bar T}}_-]^1_1
+[{\bar T}_-]^6_6[\hat{{\bar T}}_-]^2_2\Big)\no\\
&&\hspace{10mm}+[{\bar T}_-]^3_3[\hat{{\bar T}}_-]^3_3 +[{\bar
T}_-]^4_4[\hat{{\bar
T}}_-]^4_4+e^{\epsilon-\epsilon'}e^{-2\eta}\Big([{\bar
T}_-]^1_1[\hat{{\bar T}}_-]^5_5 +[{\bar T}_-]^2_2[\hat{{\bar
T}}_-]^6_6\Big)\Big\}.
 \eea
Here $[{T}_{\pm}]^\alpha_\beta$,  $[\hat {T}_{\pm}]^\alpha_\beta$, $[\bar{T}_{\pm}]^\alpha_\beta$ and
$[\hat{\bar T}_{\pm}]^\alpha_\beta$ are the operators acting
on the quantum space $V_1\otimes V_2\otimes \cdots \otimes V_N$ with the explicit expressions
\bea
&&[T_{\pm}]^{\alpha}_{\beta}=\sum_{\{\delta_i\}=1,\{\gamma_i\}=1}^4
[R^{(\pm)}_{01}]^{\alpha \gamma_1}_{\alpha_1 \delta_1}
[R^{(\pm)}_{02}]^{\alpha_1 \gamma_2}_{\alpha_2 \delta_2} \cdots
[R^{(\pm)}_{0N}]^{\alpha_{N-1} \gamma_N}_{\beta\quad\;\;\, \delta_N},\no\\
&&[\hat T_{\pm}]^{\alpha}_{\beta}=\sum_{\{\delta_i\}=1,\{\gamma_i\}=1}^4
[R^{(\pm)}_{N0}]^{\gamma_N \alpha }_{\delta_N \alpha_N}
[R^{(\pm)}_{N-10}]^{\gamma_{N-1}\alpha_{N} }_{\delta_{N-1} \alpha_{N-1}}\cdots
[R^{(\pm)}_{20}]^ {\gamma_1 \alpha_2}_{ \delta_1\beta},\no\\
&&[\bar T_{\pm}]^{\alpha}_{\beta}=\sum_{\{\delta_i\}=1,\{\gamma_i\}=1}^4[R^{(\pm)}_{\bar{0}1}]^{\alpha \gamma_1}_{\alpha_1 \delta_1} [R^{(\pm)}_{\bar{0}2}]^{\alpha_1 \gamma_2}_{\alpha_2 \delta_2}
\cdots [R^{(\pm)}_{\bar{0}N}]^{\alpha_{N-1}
\gamma_N}_{\beta\quad\;\;\, \delta_N}, \no \\
&&[\hat {\bar T}_{\pm}]^{\alpha}_{\beta}=
\sum_{\{\delta_i\}=1,\{\gamma_i\}=1}^4
[R^{(\pm)}_{N\bar{0}}]^{\gamma_N\alpha}_{\delta_N \alpha_N}
[R^{(\pm)}_{N-1\bar{0}}]^{\gamma_{N-1}\alpha_{N} }_{\delta_{N-1}
\alpha_{N-1}} \cdots [R^{(\pm)}_{1\bar{0}}]^{\gamma_1\alpha_2 }_{
\delta_1\beta}, \label{tjj-3} \eea where the repeated indicators
should be summarized, and $R^{(\pm)}_{0j}$ and
$R^{(\pm)}_{\bar{0}j}$ are the leading terms of $e^{\mp
u}R_{0j}(u)|_{u\rightarrow\pm\infty}$ and $e^{\mp
u}R_{\bar{0}j}(u)|_{u\rightarrow\pm\infty}$, respectively. The
detailed calculation shows that the eigenvalues of conserved
quantities $Q_{\pm}$ and $\bar Q_{\pm}$ can be characterized by a
quantum number $m$ (an  integer $|m|\in[0,N]$).
Then we
obtain the asymptotic behaviors of $\Lambda(u)$ and $\bar{\Lambda}(u)$ as \bea
&&\hspace{-0.8cm}\Lambda(u)|_{u\rightarrow \pm\infty}= \frac
{2}{4^{N+1}\sinh^2\eta}\Big[\cosh(\epsilon-\epsilon'-2\eta)+\cosh
(2m\eta)\Big]e^{\pm(2Nu+2u-4N\eta-4\eta)}+\cdots,\no\\
&&\hspace{-0.8cm}\bar{\Lambda}(u)|_{u\rightarrow \pm\infty}=\frac
{2}{\sinh^2 \eta}\Big[2\cosh(\epsilon-\epsilon'-2\eta)\cosh
(2m\eta)+1\Big]e^{\pm(2Nu-4N\eta)}+\cdots.\label{14Op} \eea

\section{Inhomogeneous T-Q relations}
\setcounter{equation}{0}

Acting the operator identities \eqref{Op-Product-Periodic-41}
on a common eigenstate, we obtain the
functional relations among the eigenvalues $\Lambda(u)$ and
$\bar{\Lambda}(u)$ as \bea &&
\hspace{-0.6cm}\Lambda(\pm\theta_j)\Lambda(\pm\theta_j+4\eta+i\pi)=
\frac{\sinh(\pm2\theta_j-2\eta)\sinh(\pm2\theta_j+2\eta)\sinh(\pm\theta_j-4\eta)
\sinh(\pm\theta_j+4\eta) }
{4\sinh^4\eta\cosh(\pm\theta_j-2\eta)\cosh(\pm\theta_j+2\eta)}\no\\
&&\hspace{-0.6cm}\qquad\qquad\times \prod_{l=1}^N
a(\pm\theta_j-\theta_l)c(\pm\theta_j-\theta_l+4\eta+i\pi)
a(\pm\theta_j+\theta_l)c(\pm\theta_j+\theta_l+4\eta+i\pi),\no  \\
&&\hspace{-0.6cm}
\Lambda(\pm\theta_j)\Lambda(\pm\theta_j+2\eta)=\frac{\sinh^2(\pm2\theta_j-2\eta)
\sinh(\pm\theta_j+2\eta)\sinh(\pm\theta_j-4\eta) }
{4\sinh^2\eta\cosh(\pm\theta_j)\cosh(\pm\theta_j-2\eta)
}\no\\
&&\hspace{-0.6cm}\qquad\qquad\times\prod_{l=1}^N
\tilde{\rho}_0(\pm\theta_j-\theta_l)\tilde{\rho}_0(\pm\theta_j+\theta_l)
 \bar{\Lambda}(\pm\theta_j+\eta),\no \\
 &&\hspace{-0.6cm}\Lambda(\pm\theta_j)\,\bar{\Lambda}(\pm\theta_j+3\eta)=
 \frac{2\cosh(\pm\theta_j)\cosh(\pm2\theta_j-2\eta)
\sinh(\pm\theta_j-4\eta)}
{\sinh^2\eta\sinh(\pm2\theta_j+2\eta)\sinh(\pm2\theta_j-4\eta)}\no\\
&&\hspace{-0.6cm}\qquad\qquad\times\prod_{l=1}^N
\tilde{\rho}_1(\pm\theta_j-\theta_l)\tilde{\rho}_1(\pm\theta_j+\theta_l)
 \Lambda(\pm\theta_j+2\eta+i\pi), \quad j=1,\cdots, N. \label{Op-e4} \eea
According to Eqs.(\ref{2Op}) and (\ref{1Op}), we also have other
seven constraints of $\Lambda(u)$ and
$\bar{\Lambda}(u)$ as \bea&&
\Lambda(0)=4\cosh^2\eta\prod_{j=1}^N\rho_1
(-\theta_j),\quad \Lambda(4\eta+i\pi)=4\cosh^2\eta\prod_{j=1}^N\rho_1
(-\theta_j),\label{op-relation}\\
&&\Lambda(2\eta)=\frac{\cosh^2 \eta}{2\cosh 2\eta(1+2\cosh
2\eta)}\bar{\Lambda}(\eta).\label{4Op}  \eea

The $6N$ function relations (\ref{Op-e4}) and $7$ constraints (\ref{14Op}) and (\ref{op-relation})-(\ref{4Op}) allow us sufficient information to determine the values of $\Lambda(u)$ and $\bar{\Lambda}(u)$,
which can be  expressed in terms of inhomogeneous $T-Q$ relations as
\bea
&&\Lambda(u)=\frac{\sinh(u-4\eta)\sinh(2u-2\eta)}{2\sinh^2\eta\cosh(u-2\eta)}\prod_{j=1}^N
a(u-\theta_j)a(u+\theta_j)\frac{Q^{(1)}(u+2\eta)}{Q^{(1)}(u)}\no\\[4pt]
&&\qquad+\frac{\sinh(u)\sinh(u-4\eta)}{\sinh^2\eta\sinh(2u-4\eta)
}\prod_{j=1}^Nb(u-\theta_j)b(u+\theta_j)\Big[{\sinh(2u-2\eta)
} \no\\
&&\qquad \left.\times\frac{Q^{(1)}(u-2\eta)Q^{(2)}(u+2\eta)}
{Q^{(1)}(u)Q^{(2)}(u)} +{\sinh(2u-6\eta)
}\frac{Q^{(1)}(u-i\pi)Q^{(2)}(u-2\eta)}
{Q^{(1)}(u-2\eta-i\pi)Q^{(2)}(u)}\right]\no\\
&&\qquad
+\frac{\sinh(u)\sinh(2u-6\eta)}{2\sinh^2\eta\cosh(u-2\eta)}\prod_{j=1}^N
c(u-\theta_j)c(u+\theta_j)
\frac{Q^{(1)}(u-4\eta-i\pi)}{Q^{(1)}(u-2\eta-i\pi)}\no\\[4pt]
&&\qquad+ h\prod_{j=1}^Nb(u-\theta_j)b(u+\theta_j) \frac{\sinh
u\sinh(u-4\eta)}{\sinh^2\eta}
 \frac{Q^{(1)}(u-2\eta)Q^{(1)}(u-i\pi)}{Q^{(2)}(u)},\label{tannn1} \\
&&\bar{\Lambda}(u)=\frac{1}{\sinh^2\eta}\Big\{\prod_{j=1}^N
a_1(u-\theta_j)a_1(u+\theta_j)\frac{\sinh(u-3\eta)}{\sinh(u-\eta)}\no\\
&&\qquad \times
\left[\frac{\sinh(2u)}{\sinh(2u-4\eta)}\frac{Q^{(2)}(u+3\eta)}{Q^{(2)}(u+\eta)}+
\frac{Q^{(1)}(u+\eta)Q^{(1)}(u+\eta-i\pi)Q^{(2)}(u-\eta)}{Q^{(1)}(u-\eta)
Q^{(1)}(u-\eta-i\pi)Q^{(2)}(u+\eta)}\right]\no\\
&&\qquad +\prod_{j=1}^N
b_1(u-\theta_j)b_1(u+\theta_j)\frac{\sinh(u-\eta)}{\sinh(u-3\eta)}\no\\
&&\qquad \times\left[\frac{\sinh(2u-8\eta)}{\sinh(2u-4\eta)}
\frac{Q^{(2)}(u-3\eta)}{Q^{(2)}(u-\eta)}+
\frac{Q^{(1)}(u-3\eta)Q^{(1)}(u-3\eta-i\pi)Q^{(2)}(u+\eta)}
{Q^{(1)}(u-\eta)Q^{(1)}(u-\eta-i\pi)Q^{(2)}(u-\eta)}\right]\no\\
&&\qquad +\prod_{j=1}^N
c_1(u-\theta_j)c_1(u+\theta_j)\frac{Q^{(1)}(u+\eta)Q^{(1)}(u-3\eta-i\pi)}
{Q^{(1)}(u-\eta)Q^{(1)}(u-\eta-i\pi)}\no\\
&&\qquad +\prod_{j=1}^N
c_1(u-\theta_j-i\pi)c_1(u+\theta_j-i\pi)\frac{Q^{(1)}(u+\eta-i\pi)Q^{(1)}(u-3\eta)}
{Q^{(1)}(u-\eta)Q^{(1)}(u-\eta-i\pi)}\no\\
&&\qquad +h\,
\frac{\cosh(u-\eta)\sinh(u-3\eta)}{\sinh(2u-4\eta)}\prod_{j=1}^Na_1(u-\theta_j)a_1(u+\theta_j)
\no\\
&&\qquad \times
\frac{Q^{(1)}(u+\eta)Q^{(1)}(u+\eta-i\pi)}{Q^{(2)}(u+\eta)}+h\,
\frac{\cosh(u-3\eta)\sinh(u-\eta)}{\sinh(2u-4\eta)}\no\\
&&\qquad \times \prod_{j=1}^Nb_1(u-\theta_j)b_1(u+\theta_j)
\frac{Q^{(1)}(u-3\eta)Q^{(1)}(u-3\eta-i\pi)}{Q^{(2)}(u-\eta)}
\Big\},\label{tannn3} \eea where $h$ is a parameter to be determined later (see (\ref{h-parameter}) below), the $Q$-functions are
\bea
&&Q^{(1)}(u)=\prod_{l=1}^{L_1}\sinh\frac 12(u-\mu_l^{(1)}-\eta)\sinh\frac 12(u+\mu_l^{(1)}-\eta),\no\\
&&Q^{(2)}(u)=\prod_{k=1}^{L_2}\sinh(u-\mu_k^{(2)}-2\eta)\sinh(u+\mu_k^{(2)}-2\eta),\no\\
&& a_1(u)=2\sinh(u-3\eta), \quad b_1(u)=2\sinh(u-\eta),\no\\&&
c_1(u)=4\sinh\frac 12(u-3\eta)\cosh\frac
12(u-\eta).\label{a1b1c1}
\eea
$L_1$ is the number of Bethe roots $\{\mu_l^{(1)}\}$ and $L_2$ is
the number of Bethe roots $\{\mu_k^{(2)}\}$. The regularities of
eigenvalues $\Lambda(u)$ and $\bar{\Lambda}(u)$ require that the
Bethe roots $\{\mu^{(1)}_l\}$ and $\{\mu^{(2)}_k\}$ satisfy the
Bethe ansatz equations (BAEs) \bea
&&\frac{Q^{(1)}(\mu_l^{(1)}+3\eta)Q^{(2)}(\mu_l^{(1)}+\eta)}
{Q^{(1)}(\mu_l^{(1)}-\eta)Q^{(2)}(\mu_l^{(1)}+3\eta)}
=-\frac{\sinh(\mu_l^{(1)}+\eta)}
{\sinh(\mu_l^{(1)}-\eta)}\no\\
&&\qquad\qquad
\times\prod_{j=1}^N\frac{\sinh\frac12(\mu_l^{(1)}+\eta-\theta_j)
\sinh\frac12(\mu_l^{(1)}+\eta+\theta_j)}{\sinh\frac12(\mu_l^{(1)}-\eta-\theta_j)
\sinh\frac12(\mu_l^{(1)}-\eta+\theta_j)},\label{BAEs-2231} \quad
l=1,\cdots,
L_1,\\
&&\frac{\sinh(2\mu_k^{(2)}+2\eta)}{\sinh(2\mu_k^{(2)})}
\frac{Q^{(2)}(\mu_k^{(2)}+4\eta)}{Q^{(1)}(\mu_k^{(2)}+2\eta)Q^{(1)}(\mu_k^{(2)}+2\eta-i\pi)}\no\\
&&\qquad\qquad
+\frac{\sinh(2\mu_k^{(2)}-2\eta)}{\sinh(2\mu_k^{(2)})}
\frac{Q^{(2)}(\mu_k^{(2)})}{Q^{(1)}(\mu_k^{(2)})Q^{(1)}(\mu_k^{(2)}-i\pi)}
=-h, \quad k=1,\cdots, L_2.   \label{BAEs-223} \eea From the
degree analysis of polynomials $\Lambda(u)$ and
$\bar{\Lambda}(u)$, we obtain that the numbers of Bethe roots
$\{\mu^{(1)}_l\}$ and $\{\mu^{(2)}_k\}$ should be the same, i.e.,
$L_1=L_2$. According to the asymptotic behaviors of $\Lambda(u)$
and $\bar{\Lambda}(u)$, the value of $h$ is determined as \bea
h=(-1)^{L_1}4^{L_1}\Big\{2\cosh(\epsilon-\epsilon'-2\eta)-2\cosh[2(L_1+1)\eta]\Big\}.\label{h-parameter}
\eea Meanwhile, the quantum number $m$ is related with the number
of Bethe roots as $m=L_1-N$\footnote{If $m\leq 0$, we have $0\leq
L_1\leq N$. When $m\geq 0$, we have $N\leq L_1\leq 2N$.}.

Here we present the numerical solutions of the $T-Q$ relation (\ref{tannn1}) with BAEs (\ref{BAEs-2231})-(\ref{BAEs-223}) for the $N=2$ case in Table \ref{table_num}.
The eigenvalue calculated from (\ref{tannn1}) is the same as that from the exact diagonalization of the transfer matrix $t(u)$ (\ref{trweweu1110}) with open boundary conditions.
Numerical solutions with random choice of $u$
and $\eta$ for some small size imply that the solution (\ref{tannn1}) indeed gives the complete solutions of
the model.
We further remark that for $L=0$, the $T-Q$ relation will give the same eigenvalue as the $L=4$ in the Table.

\begin{landscape}

\begin{table}
\vspace{-8mm}
\caption{Solutions of BAEs (\ref{BAEs-2231})-(\ref{BAEs-223}) for the $T-Q$ relation (\ref{tannn1}), $N=2$, $u=0.2$, $\eta=0.4$, $\{\theta_j\}=0$,  $\epsilon=\epsilon'=0$. The numbers of the two sets of Bethe roots are the same $L=L_1=L_2$. The symbol $n$ indicates the number of the spectrum $\Lambda(u)$. }\label{table_num}
\begin{tabular}{|c|c|c|c|c|c|c|}
\hline\hline $\mu^{(1)}_1$ & $\mu^{(1)}_2$ & $\mu^{(1)}_3$ & $\mu^{(1)}_4$ & $\mu^{(2)}_1$ & $\mu^{(2)}_2$  \\ \hline
$2.8213-7.3824i$ & $2.8566-11.7923i$  &  $2.6075-9.6328i$  & $-15.4519+9.5555i$  & $5.7360-20.1799i$ & $-15.9439+3.2723i$ \\
$-1.8106-5.5833i$ & $-0.6073-3.3582i$ &  $-9.6710-14.6796i$  & $-1.7855-0.7885i$ & $8.1575-2.9733i$ & $-10.2358-11.4398i$ \\
$-1.1119-3.1416i$ &   ------         &         ------        &      ------        &  $1.7606+0.0000i$  &    ------        \\
$-1.1119+3.1416i$ &$-12.7134+2.2785i$ & $-12.7134-4.0046i$ &      ------        &  $13.5955+4.0046i$  &  $1.7606-3.1416i$  \\

$-12.8331+0.7576i$ &  $-0.0000-3.1416i$  &    ------        &      ------        & $20.7772-1.5018i$  &  $20.9923-0.1902i$ \\
$-0.7650+1.2091i$  &  $-0.7650-1.2091i$  &    ------        &      ------        & $10.3808+0.9039i$  &  $1.0705-1.5708i$ \\
$0.0751-0.3828i$   &     ------        &      ------        &      ------        &  $-0.7428-2.6351i$ &  ------ \\
$8.5971+2.8481i$  &  $0.0751-0.3828i$  &  $-8.5971+0.2935i$ &      ------        &  $-9.1416+0.6080i$ & $-0.7428+6.7896i$ \\

$-0.0751-0.3828i$  &     ------        &      ------        &      ------        &  $0.7428+0.5064i$  &  ------  \\
$-0.0751+5.9004i$  & $8.8382+6.7463i$  & $-8.8382+8.9616i$  &      ------        &  $-9.3827+2.9929i$ & $-0.7428+12.0599i$ \\

$0.0051+0.4591i$  &  $-0.0676+1.4934i$ &      ------        &      ------        &  $9.3206-0.5054i$  & $-9.0327+0.6314i$ \\
$0.1816-1.3704i$  &  $-0.0175-0.4666i$ &      ------        &      ------        &  $-9.1912-0.4531i$ & $-9.3950+2.4425i$ \\
$0.0000-0.0000i$  & $-14.0480+2.8382i$ &      ------        &      ------        & $22.9513-2.7003i$  & $22.9864-1.7707i$ \\

$-0.0393+0.0000i$ &     ------        &       ------        &      ------        &  $0.4925-0.0000i$ &  ------  \\
$0.0393+0.0000i$  & $8.6900-0.0624i$  &  $8.6901-3.2040i$   &      ------        &  $-9.2346+2.8895i$ &  $9.2346-0.3770i$\\
$-0.0000-1.6458i$ &  $-0.0000+0.0049i$ &      ------        &      ------        & $-0.4919-0.0000i$  & $7.3047-1.4074i$ \\

\hline\hline $\mu^{(2)}_3$ & $\mu^{(2)}_4$    & $\Lambda(u)$      &      $L$ & $n$  &   \\ \hline
$-5.7344+12.8484i$  &  $-5.7362+5.5171i$ & $0.6400-0.0000i$   & $4$ & $1$  &  \\
$9.4903+3.1677i$  &  $-8.0876-4.9070i$ &   $0.6400-0.0000i$   & $4$ & $2$  &  \\
    ------        &      ------        & $0.6884-0.0000i$   & $1$  & $3$  &   \\
$-12.9204+5.4201i$&      ------        & $0.6884-0.0000i$   & $3$  & $4$   & \\

    ------        &      ------        & $0.7049-0.0000i$   & $2$   & $5$  &   \\
    ------        &      ------        & $0.7782-0.0000i$   & $2$   & $6$  &   \\
    ------        &      ------        & $1.7412-0.3231i$   & $1$   & $7$  &   \\
$9.1416+3.1626i$  &      ------        & $1.7412-0.3231i$   & $3$   & $8$  &   \\

    ------        &      ------        & $1.7412+0.3231i$   & $1$   & $9$  &   \\
$-9.3827+5.5055i$ &      ------        & $1.7412+0.3231i$   & $3$   & $10$  &   \\

    ------        &      ------        & $1.7907-0.0000i$   & $2$    & $11$  &   \\
    ------        &      ------        & $1.7907-0.0000i$   & $2$    & $12$  &   \\
    ------        &      ------        & $6.0819-0.0000i$   & $2$    & $13$  &    \\

    ------        &      ------        & $6.2595-0.0000i$   & $1$    & $14$  &    \\
$0.4925-3.1416i$  &      ------        & $6.2595-0.0000i$   & $3$    & $15$  &    \\

    ------        &      ------        & $6.9792-0.0000i$   & $2$    & $16$  &    \\

 \hline\hline\end{tabular}
\end{table}
\end{landscape}

We shall give some remarks about the obtained eigenvalues. The BAEs (\ref{BAEs-2231}) are
homogeneous while the BAEs (\ref{BAEs-223}) are inhomogeneous. This is because that the
non-diagonal boundary reflections break the $U(1)$ symmetry of the system.
Due to the present form of reflection matrix \eqref{K-matrix-1}, which is diagonal in a $2\times 2$ subspace,
the Bethe roots in this subspace satisfy the homogeneous BAEs.
The existence of one sets of homogeneous BAEs consists with the fact
that there is only one good quantum number $m$.
Another things we should mentioned is that during the construction of $T-Q$ relation, the BAEs obtained from the regularities of $\Lambda(u)$ and $\bar{\Lambda}(u)$
should be the same.  It is easy to check that $\Lambda(u)$ and $\bar{\Lambda}(u)$ satisfy the
functional relations (\ref{Op-e4}) and the additional constraints (\ref{4Op}). Therefore, we conclude that
the analytical expressions \eqref{tannn1} and \eqref{tannn3} are the
eigenvalues of the transfer matrices $t(u)$ and
$\bar{t}(u)$, respectively. It is noted that the eigenvalues and associated BAEs have the well-defined homogeneous limit.

Based on the exact solution (\ref{tannn1}) and (\ref{BAEs-2231})-(\ref{BAEs-223}) of $\Lambda(u)$, we can obtain the energy spectrum of the Hamiltonian \eqref{hh}
\begin{eqnarray}
E= \frac{\partial \ln \Lambda(u)}{\partial
u}|_{u=0,\{\theta_j\}=0}.
\end{eqnarray}

\section{Periodic boundary condition case}
\setcounter{equation}{0}

We shall note that the above method is universal, which is also valid for the quantum integrable systems with $U(1)$ symmetry.
For this purpose, we consider the exact solution of the $A_3^{(2)}$ model with the periodic boundary condition.

In the periodic case, the transfer matrix $t^{(p)}(u)$ and the fused one $\bar{t}^{(p)}(u)$ are defined as
\bea t^{(p)}(u)=tr_0 T_0(u), \quad  \bar{t}^{(p)}(u)=tr_{\bar{0}} T_{\bar{0}}(u), \label{1117-1}\eea
where the monodromy matrices $T_0(u)$ and $T_{\bar 0}(u)$ are given by (\ref{Mon-1}) and \eqref{M2on1-2}, respectively.
From the Yang-Baxter equations \eqref{20190802-1} and (\ref{201908102-1}), we can prove
that the transfer matrices $ t^{(p)}(u)$ and $ \bar{t}^{(p)}(u)$
satisfy the commutation relations \bea [t^{(p)}(u),
{t}^{(p)}(v)]=[t^{(p)}(u),
\bar{t}^{(p)}(v)]=0. \eea
Taking the partial trace of Eq.(\ref{98opr-11}) in the auxiliary
spaces and using the relation (\ref{98opr-1}), we obtain the
operator product identities \bea &&
t^{(p)}(\theta_j)\,t^{(p)}(\theta_j+4\eta+i\pi)=\prod_{l=1}^N
a(\theta_j-\theta_l)c(\theta_j-\theta_l+4\eta+i\pi)\times {\rm id}, \no  \\
&& t^{(p)}(\theta_j)\,t^{(p)}(\theta_j+2\eta)= \prod_{l=1}^N
\tilde{\rho}_0(\theta_j-\theta_l)\,\bar{t}^{(p)}(\theta_j+\eta),\no  \\
&&t^{(p)}(\theta_j)\,\bar{t}^{(p)}(\theta_j+3\eta)=\prod_{l=1}^N
\tilde{\rho}_1(\theta_j-\theta_l)\,t^{(p)}(\theta_j+2\eta+i\pi),\quad
j=1,\cdots,N.\label{fut1p-5}\eea
From the definitions \eqref{1117-1}, we obtain the asymptotic behaviors of fused transfer
matrices
\bea && t^{ (p)}(u)|_{u\rightarrow \pm\infty}=e^{\pm
(Nu-\sum_{j=1}^N\theta_j)}
\sum_{\alpha=1}^4[T_{\pm}]^{\alpha}_{\alpha}+\cdots,\no \\
&& \bar{t}^{(p)}(u)|_{u\rightarrow \pm\infty}=e^{\pm(
Nu-\sum_{j=1}^N\theta_j)} \sum_{\alpha=1}^6[\bar
T_{\pm}]^{\alpha}_{\alpha} +\cdots, \label{fuwwwtpl-7} \eea where
$\sum_{\alpha=1}^4[T_{\pm}]^{\alpha}_{\alpha}$ and
$\sum_{\alpha=1}^6[\bar T_{\pm}]^{\alpha}_{\alpha}$ are the
conserved quantities. From the direct calculation, we find that
the eigenvalues of  $\sum_{\alpha=1}^4[T_{\pm}]^{\alpha}_{\alpha}$
and $\sum_{\alpha=1}^6[\bar{T}_{\pm}]^{\alpha}_{\alpha}$ can be
quantified by two quantum numbers $m_1$ and $m_2$ as
$2^{1-N}[\cosh(m_1\eta)+\cosh(m_2\eta)]e^{\mp 2N\eta}$ and
 $2\{1+\cosh[(m_1+m_2)\eta]
+\cosh[(m_1-m_2)\eta]\}e^{\mp 2N\eta}$, respectively, where
$m_1\in [0,N]$ and $0\leq |m_2| \leq N-m_1$. Then the asymptotic
behaviors of transfer matrices $t^{(p)}(u)$ and $\bar{t}^{(p)}(u)$
on some subspace which can be parameterized by two integers $m_1$
and $m_2$  read \bea &&\hspace{-1.4cm} t^{ (p)}(u)|_{u\rightarrow
\pm\infty}= 2^{1-N}[\cosh(m_1\eta)
+\cosh(m_2\eta)]e^{\pm (Nu-\sum_{j=1}^N\theta_j-2N\eta)} +\cdots,\no \\
&&\hspace{-1.4cm} \bar{t}^{(p)}(u)|_{u\rightarrow \pm\infty}=
2\{1+\cosh[(m_1+m_2)\eta] +\cosh[(m_1-m_2)\eta]\} e^{\pm
(Nu-\sum_{j=1}^N\theta_j-2N\eta)} +\cdots.\label{fuwwsdsdw2tpl-7}
\eea

Suppose the eigenvalues of $t^{(p)}(u)$ and $\bar{t}^{(p)}(u)$ as $\Lambda^{(p)}(u)$ and
$\bar{\Lambda}^{(p)}(u)$, respectively.
From Eqs.(\ref{fut1p-5}) and \eqref{fuwwsdsdw2tpl-7}, we obtain that $\Lambda^{(p)}(u)$ and
$\bar{\Lambda}^{(p)}(u)$ should satisfy the constraints
\bea &&
\Lambda^{(p)}(\theta_j)\,\Lambda^{(p)}(\theta_j+4\eta+i\pi)=\prod_{l=1}^N
a(\theta_j-\theta_l)c(\theta_j-\theta_l+4\eta+i\pi),\no  \\
&& \Lambda^{(p)}(\theta_j)\,\Lambda^{(p)}(\theta_j+2\eta)=
\prod_{l=1}^N
\tilde{\rho}_0(\theta_j-\theta_l)\,\bar{\Lambda}^{(p)}(\theta_j+\eta),\no  \\
&&\Lambda^{(p)}(\theta_j)\bar{\Lambda}^{(p)}(\theta_j+3\eta)=\prod_{l=1}^N
\tilde{\rho}_1(\theta_j-\theta_l)\,\Lambda^{(p)}(\theta_j+2\eta+i\pi),\quad
j=1,\cdots,N, \no \\
&& \Lambda^{ (p)}(u)|_{u\rightarrow \pm\infty}=
2^{1-N}[\cosh(m_1\eta)
+\cosh(m_2\eta)]e^{\pm (Nu-\sum_{j=1}^N\theta_j-2N\eta)} +\cdots,\no \\
&& \bar{\Lambda}^{(p)}(u)|_{u\rightarrow \pm\infty}=
2\{1+\cosh[(m_1+m_2)\eta]
+\cosh[(m_1-m_2)\eta]\}\no \\
&&\hspace{3cm} \times e^{\pm (Nu-\sum_{j=1}^N\theta_j-2N\eta)}
+\cdots.\label{fuwww2tpl-7} \eea

The eigenvalues $\Lambda^{(p)}(u)$ (resp. $\bar{\Lambda}^{(p)}(u)$)
is a polynomial of $e^u$ with degree $2N$ (a polynomial of $e^{2u}$ with degree  $N$ respectively) up to an overall factor $e^{-Nu}$.
Therefore, $\Lambda^{(p)}(u)$ and $\bar{\Lambda}^{(p)}(u)$
can be completely determined by at least $3N+2$ constraints.
Then we arrive at that $3N$ functional relations together with $4$ asymptotic behaviors \eqref{fuwww2tpl-7}  can determine the eigenvalues $\Lambda^{(p)}(u)$ and $\bar{\Lambda}^{(p)}(u)$, which are
expressed by the $T-Q$ relations as \bea \Lambda^{(p)}(u)&=&\prod_{j=1}^N
a(u-\theta_j)\frac{Q_{p}^{(1)}(u+2\eta)}{Q_{p}^{(1)}(u)}
+\prod_{j=1}^Nb(u-\theta_j)\Big\{ \frac{Q_{p}^{(1)}(u-2\eta)
Q_{p}^{(2)}(u+2\eta)}{Q_{p}^{(1)}(u)Q_{p}^{(2)}(u)}\no\\
&&+\frac{Q_{p}^{(1)}(u-i\pi)Q_{p}^{(2)}(u-2\eta)}{Q_{p}^{(1)}(u-2\eta-i\pi)Q_{p}^{(2)}(u)}\Big\}
+\prod_{j=1}^N c(u-\theta_j)\,
\frac{Q_{p}^{(1)}(u-4\eta-i\pi)}{Q_{p}^{(1)}(u-2\eta-i\pi)}, \label{T-Q-Periodic1}\\
\bar{\Lambda}^{(p)}(u)&=&\prod_{j=1}^N
a_1(u-\theta_j)\left[\frac{Q^{(2)}_p(u+3\eta)}{Q^{(2)}_p(u+\eta)}
+\frac{Q^{(1)}_p(u+\eta)Q^{(1)}_p(u+\eta-i\pi)Q^{(2)}_p(u-\eta)}
{Q^{(1)}_p(u-\eta)Q^{(1)}_p(u-\eta-i\pi)Q^{(2)}_p(u+\eta)}\right]\no\\
&&+\prod_{j=1}^N
b_1(u-\theta_j)\left[\frac{Q^{(2)}_p(u-3\eta)}{Q^{(2)}_p(u-\eta)}
+\frac{Q^{(1)}_p(u-3\eta)Q^{(1)}_p(u-3\eta-i\pi)Q^{(2)}_p(u+\eta)}
{Q^{(1)}_p(u-\eta)Q^{(1)}_p(u-\eta-i\pi)Q^{(2)}_p(u-\eta)}\right]\no\\
&&+\prod_{j=1}^N
c_1(u-\theta_j)\frac{Q^{(1)}_p(u+\eta)Q^{(1)}_p(u-3\eta-i\pi)}
{Q^{(1)}_p(u-\eta)Q^{(1)}_p(u-\eta-i\pi)}\no\\
&&+\prod_{j=1}^N
c_1(u-\theta_j-i\pi)\frac{Q^{(1)}_p(u+\eta-i\pi)Q^{(1)}_p(u-3\eta)}
{Q^{(1)}_p(u-\eta)Q^{(1)}_p(u-\eta-i\pi)}, \label{T-Q-Periodic2}
 \eea where the definition of the functions $a_1(u)$, $b_1(u)$, and $c_1(u)$ is in (\ref{a1b1c1}), and
\bea
&&Q_{p}^{(1)}(u)=\prod_{l=1}^{L_1}\sinh\frac
12(u-\mu_l^{(1)}-\eta),\quad
Q_{p}^{(2)}(u)=\prod_{k=1}^{L_2}\sinh(u-\mu_k^{(2)}-2\eta). \eea

The regularity analyses of the $T-Q$ relations
(\ref{T-Q-Periodic1})-(\ref{T-Q-Periodic2}) lead to that the Bethe
roots $\{\mu^{(1)}_l\}$ and $\{\mu^{(2)}_k\}$ should satisfy the
BAEs \bea &&
\frac{Q_{p}^{(1)}(\mu_l^{(1)}+3\eta)Q_{p}^{(2)}(\mu_l^{(1)}+\eta)}
{Q_{p}^{(1)}(\mu_l^{(1)}-\eta)Q_{p}^{(2)}(\mu_l^{(1)}+3\eta)}
=-\prod_{j=1}^N \frac{\sinh\frac 12(\mu_l^{(1)}+\eta-\theta_j)}{\sinh\frac 12(\mu_l^{(1)}-\eta-\theta_j)},\;\;l=1,\cdots, L_1, \label{BAEs-31} \\[6pt]
&&\frac{Q_{p}^{(1)}(\mu_k^{(2)})Q_{p}^{(1)}(\mu_k^{(2)}-i\pi)Q_{p}^{(2)}(\mu_k^{(2)}+4\eta)}
{Q_{p}^{(1)}(\mu_k^{(2)}+2\eta)Q_{p}^{(1)}(\mu_k^{(2)}+2\eta-i\pi)Q_{p}^{(2)}(\mu_k^{(2)})}
=-1, \quad k=1,\cdots, L_2,\label{BAEs-3} \eea where $L_1\leq N$
and $L_2\leq L_1$. We shall note that the BAEs (\ref{BAEs-31})
and (\ref{BAEs-3}) are homogeneous, because the
periodic boundary condition does not break the $U(1)$ symmetry. The quantum numbers $m_1$ and $m_2$ characterizing the
conserved quantities
$\sum_{\alpha=1}^4[T_{\pm}]^{\alpha}_{\alpha}$ and
$\sum_{\alpha=1}^6[\bar T_{\pm}]^{\alpha}_{\alpha}$ are
related with the numbers of Bethe roots as
\begin{eqnarray}
m_1=N-L_1, \quad m_2=L_1-2L_2. \label{m}
\end{eqnarray}
The eigenvalues (\ref{T-Q-Periodic1})-(\ref{T-Q-Periodic2}) and
associated BAEs (\ref{BAEs-31})-(\ref{BAEs-3}) have the
well-defined homogeneous limit. These results with the constraint
$\{\theta_j\}=0$ are coincide with the previous ones obtained by using the functional or nested algebraic Bethe ansatz
\cite{NYRes, NYReshetikhin2}.

\section{Discussion}

In this paper, we have studied the exact solution of quantum integrable model
associated with the $A^{(2)}_3$ twisted Lie algebra. We give a detailed analysis of the
fusion properties, including  the open chain and the periodic one.
We obtain the closed recursive fusion relations and
additional constraints among the fused transfer matrices.
Based on them and with the help of polynomials analysis, we obtain the eigenspectrum and related Bethe ansatz equations of the system.
The results provided in this paper can be generalized to the $A^{(2)}_n$ model with arbitrary $n$ and integrable models with the other twisted Lie algebras.

\section*{Acknowledgments}

We would like to thank Professor Y. Wang for his valuable discussions and continuous encouragement.
The financial supports from National Key R\&D Program of China (Grant No. 2021YFA1402104), the National Natural Science Foundation of China (Grant Nos. 12175180, 12105221, 12074410, 12047502, 12075177, 11934015,
11975183, 11947301, 91536115, 12275214, 12205235 and 12105221), Major Basic Research Program of Natural Science of
Shaanxi Province (Grant Nos. 2021JCW-19, 2017KCT-12 and 2017ZDJC-32), the Scientific Research Program Funded by Shaanxi Provincial Education Department (Grant No. 21JK0946), Australian Research Council (Grant No. DP 190101529), Strategic
Priority Research Program of the Chinese Academy of Sciences (Grant No. XDB33000000), Beijing National Laboratory for Condensed Matter Physics (Grant No. 202162100001), Shaanxi Province Key Laboratory
of Quantum Information and Quantum Optoelectronic Devices, Xi'an
Jiaotong University, and the Double First-Class University Construction Project of Northwest
University are gratefully acknowledged.

\section*{Appendix A. Expression of the $R$-matrix $R_{\bar{1}2}(u)$}
\setcounter{equation}{0}
\renewcommand{\theequation}{A.\arabic{equation}}
In this appendix, we  give the explicit expression of the $R$-matrix $R_{\bar{1}2}(u)$ define in (\ref{uf-12}) as
\begingroup
\renewcommand*{\arraystretch}{0.1}
\begin{equation}
   R_{\bar{1}2}(u)= \begin{pmatrix}\setlength{\arraycolsep}{0.5pt}
    \begin{array}{cccc|cccc|cccc|cccc|cccc|cccc}
    r_1&&&&& &&&&&& &&&&&& &&&&&&  \\
    &r_1&&&& &&&&&& &&&&&& &&&&&& \\
    &&r_2& &&r_7 && &r_8&&& &r_{10}&&&&& &&&&&&  \\
    &&&r_{2} && && &&r_9&& &&r_{11}&& &r_{12}& &&&&&&  \\
    \hline&&&&r_{1}& &&&&&& &&&&&& &&&&&&  \\
    &&r_7& &&r_{2} && &r_8&&& &-r_{10}&&&&& &&&&&& \\
    &&& && &r_{1}&&&&& &&&&&& &&&&&&  \\
    &&& && &&r_{2} & &&r_9& &&&-r_{11}& && && &r_{12}&&& \\
    \hline&&r_{13}& &r_{13}& &&&r_{3}&&& &&&&&& &&&&&& \\
    &&&r_{14} && && &&r_{4}&& &&r_{15}&& &r_9& &&&&&&  \\
     &&& && &&r_{14} &&&r_{4}& &&&-r_{15}& && && &r_{9}&&& \\
     &&& && && &&&&r_{3} &&&& && &r_8& &&r_8&&  \\
   \hline  &&r_{16}& &-r_{16}& & & &&&& &r_{5}&&&&& &&&&&&  \\
    &&&r_{17} && && &&-r_{15}&& &&r_{6}&& &-r_{11}& &&&&&&  \\
   &&& && &&-r_{17} &&&r_{15}& &&&r_{6}& && &&&r_{11}&&&  \\
   &&&&& &&&&&& &&&&r_{5}&& &-r_{10}& &&r_{10}&& \\
     \hline&&&r_{18} && && &&r_{14}&& &&-r_{17}&&&r_{2}& &&&&&&  \\
      &&&&& &&&&&& &&&&&&r_{1} &&&&&&  \\
    &&& && && &&&&r_{13} &&&&-r_{16}&& &r_{2}& &&r_7&&  \\
    &&&&& &&&&&& &&&&&& &&r_{1}&&&&  \\
    \hline&&& &&  &&r_{18} &&&r_{14}& &&&r_{17}&&& &&&r_{2}&&& \\
     &&& && && &&&&r_{13} &&&&r_{16}&& &r_7&&&r_{2}&& \\
     &&&&& &&&&&& &&&&&& &&&&&r_{1}&  \\
      &&&&& &&&&&& &&&&&& &&&&&&r_{1}  \\
\end{array}
    \end{pmatrix},   \label{rsp}
\end{equation}\noindent
\endgroup
\bea&&r_1=2\sinh(u-3\eta),\ r_2=2\sinh(u-\eta),\
r_3=4\sinh\frac12(u-3\eta)\cosh\frac12(u-\eta),\no\\
&&r_4=2(\sinh(u-2\eta)+\sinh2\eta\sinh\eta),\
r_5=4\sinh\frac12(u-\eta)\cosh\frac12(u-3\eta),\no\\
&&r_6=2(\sinh(u-2\eta)-\sinh2\eta\sinh\eta),\ r_7=-2\sinh
2\eta,\no\\
&&r_8=-4e^{-\frac
u2}\sinh\eta\sqrt{\cosh\eta}\sinh\frac12(u-3\eta),\
r_9=-4e^{-\frac
u2+\eta}\sinh\eta\sqrt{\cosh\eta}\cosh\frac12(u-\eta),\no\\
&&r_{10}=4e^{-\frac
u2}\sinh\eta\sqrt{\cosh\eta}\cosh\frac12(u-3\eta),\
r_{11}=4e^{-\frac
u2+\eta}\sinh\eta\sqrt{\cosh\eta}\sinh\frac12(u-\eta),\no\\
&&r_{12}=2e^{-\eta}\sinh2\eta,\ r_{13}=-e^ur_8,\
r_{14}=e^{u-2\eta}r_9, \  r_{15}=-2\sinh\eta\sinh2\eta,\no\\
&&r_{16}=e^ur_{10},\ r_{17}=-e^{u-2\eta}r_{11},\
r_{18}=2e^{\eta}\sinh2\eta.\label{A-2}\eea
The above expression allows us to derive the very properties (\ref{nrp}) of the resulting
fused $R$-matrix $R_{\bar{1}2}(u)$.


\end{document}